\documentclass[twocolumn,showpacs,preprintnumbers,amsmath,amssymb,prb,aps]{revtex4}

\usepackage{epsfig}
\usepackage{graphicx}
\usepackage{dcolumn}
\usepackage{bm}

\begin{document}

\title{Comment on "Electronic structure and orbital ordering of
SrRu$_{1-x}$Ti$_x$O$_3$: GGA+U investigations"}

\author{Kalobaran Maiti}
 \altaffiliation{Electronic mail: kbmaiti@tifr.res.in}

\affiliation{Department of Condensed Matter Physics and Materials'
Science, Tata Institute of Fundamental Research, Homi Bhabha Road,
Colaba, Mumbai - 400 005, INDIA}

\date{\today}

\begin{abstract}

In the paper, PRB {\bf 77}, 085118 (2008), the authors conclude that
the observation of Ti-doping induced half-metallicity in
SrRu$_{1-x}$TI$_x$O$_3$ within the limit of local density
approximations is not valid as the experimental results indicate
insulating behavior. It was described that the metal-insulator
transition (MIT) at $x$ = 0.5 observed in this system appears due to
the enhancement of on-site Coulomb repulsion strength, $U$ with Ti
substitutions, $x$. The MIT primarily depends on $U$ and partially
on $x$ and/or disorder. All these conclusions are in sharp contrast
to the experimental observations, which predicted Anderson
insulating phase at $x$ = 0.5 (finite localized density of states at
the Fermi level). The hard gap due to electron correlation appears
at much higher $x$ ($\sim$ 0.8). In addition, it is well established
that homovalent substitution has negligible influence on on-site $U$
(a local variable). These inconsistencies appear due to the fact
that the calculated results representing the bulk electronic
structure are compared with the experimental results dominated by
surface contributions. The experimental bulk spectra, already
available in the literature exhibit finite density of states at the
Fermi level even for $x\geq$~0.5 sample, which suggests that the
conclusion of half metallic phase in an earlier study is reasonable.
Thus, the insulator to metal transition in these systems is driven
by bandwidth, $W$ rather than $U$ and disorder plays dominant role
in the metal-insulator transition. One needs to consider the bulk
spectra to reproduce numerically the bulk electronic structure.

\end{abstract}


\maketitle

Ruthenates have drawn significant attention in the recent times due
to many interesting properties exhibited by these systems such as
non-Fermi liquid behavior, unusual magnetism, superconductivity etc.
SrRuO$_3$ forms in perovskite structure and is a ferromagnetic
metal. It was observed experimentally that Ti doping at the Ru sites
in SrRuO$_3$ (SrRu$_{1-x}$Ti$_x$O$_3$) thin films leads to an
interesting evolution of the electronic properties ranging from
weakly correlated metal to a band insulator via several intermediate
phases for different values of $x$.\cite{jkim,abbate,kkim}

A recent study\cite{lin} based on generalized gradient
approximations (GGA) using {\scriptsize VASP} package observed that
Ti substitution at the Ru sites in SrRuO$_3$ leads to a metal to
half-metal transition around $x$ = 0.5. The authors conclude that
such observation is in disagreement with the experimental
observation of a metal-insulator transition as a function of
Ti-dopant concentration. It was suggested that a larger $U$ would be
more reasonable in the higher dopant concentration to achieve an
insulating ground state. In addition, it was predicted that observed
metal insulator transitions in this system is primarily driven by
on-site Coulomb repulsion. Disorder and/or Ti substitution are only
partially responsible in determining the metal-insulator transition.

All the above conclusions are in sharp contrast to the experimental
results. Firstly, the experimental results\cite{kkim} based on
transport measurements on thin films suggest an Anderson insulating
phase at $x$ = 0.5. This phase correspond to finite electronic
density of states at the Fermi level, which are localized due to the
effect of disorder. {\it Both} transport and photoemission
studies\cite{jkim,kkim} do not indicate signature of hard gap at $x$
= 0.5. The occurrence of hard gap is predicted at much higher Ti
concentration ($x \sim$ 0.8).\cite{kkim} It is clear that disorder
play a significant role in determining the electronic properties in
this system. In addition, it is already well
known\cite{FujiRMP,GabiRMP} that even in strongly correlated 3$d$
transition metal oxides, the magnitude of on-site Coulomb repulsion
strength, $U$ is not significantly sensitive to the homovalent
substitutions. This is reasonable as $U$ is a local variable. Thus,
a conclusion of change in $U$ with Ti substitution is unusual.

Various photoemission studies of such ruthenates reveal that the
intensity of the correlation induced lower Hubbard band is
significantly weaker than the intensity of the coherent feature that
represents the delocalized electronic states in the vicinity of the
Fermi level.\cite{PRBR,fujimori} Thus, the electron correlation
strength is significantly weak in these systems compared to that
often found in highly correlated 3$d$ transition metal oxides. This
again rules out the possibility of Ti-substitution induced
enhancement of $U$ in such a weakly correlated electron systems
leading to insulating phase. We believe that the inconsistencies
observed here have a different origin as described below.

The results from band structure calculations represent the bulk
electronic structure. Therefore it is desirable to compare the
calculated results with the properties obtained from bulk materials
and the spectral function representing the bulk electronic structure
of the concerned systems. All the experimental results referred in
the paper are obtained from thin film samples. The photoemission
spectra provide the direct representation of the electronic density
of state. The experimental photoemission results referred in the
paper have significant surface contributions. While it is often
considered that sufficiently thick films exhibit properties close to
that expected in the bulk of the material, it is observed that the
substrate induced strain persists even in thick films and the
two-dimensional topology of these films lead to significantly
different electronic and magnetic properties.\cite{cao} Such
confinement effects are used in numerous occasions to generate
quantum wells, in-plane magnetizations {\it etc}.

In addition, experiments on various transition metal oxides reveal
significantly different surface and bulk electronic
structures.\cite{PRBR,fujimori,LCVOPRL,Y2Ir2O7} Even the samples in
the form of thin films also exhibit surface-bulk differences in the
electronic structure.\cite{fujimori} Moreover, in the present case,
it has been demonstrated\cite{PRB-Ravi} experimentally that the
surface and bulk electronic structure in SrRu$_{1-x}$Ti$_x$O$_3$ are
significantly different. Part of the results are reproduced in the
figure for a representative case. The experimental spectra were
obtained using Gammadata Scienta analyzer, SES2002 and
monochromatized photon sources. The energy resolution for $x$-ray
photoemission (XP) and He~{\scriptsize II} photoemission
measurements were set at 300~meV and 4~meV, respectively. The
details of sample preparation and characterizations are described
elsewhere.\cite{PRB-Ravi}

\begin{figure}
\vspace{-2ex}
\includegraphics[angle=0,width=0.5\textwidth]{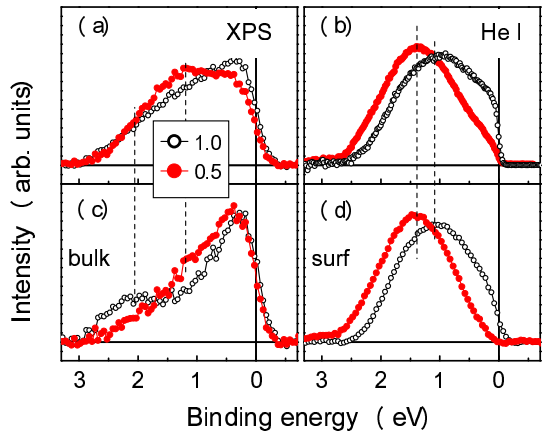}
\vspace{-36ex}
\caption{(color online) (a) $X$-ray photoemission spectra, (b) He
{\scriptsize I} spectra, (c) bulk spectra and (d) surface spectra of
SrRuO$_3$ (open circles) and SrRu$_{0.5}$Ti$_{0.5}$O$_3$ (solid
circles), respectively.}
 \vspace{-2ex}
\end{figure}

In Fig. 1(a) and 1(b), we show the $x$-ray photoemission and He
{\scriptsize I} spectra of SrRuO$_3$ (open circles) and
SrRu$_{0.5}$Ti$_{0.5}$O$_3$ (solid circles). The spectral region
shown here represents the density of states having primarily Ru 4$d$
character as also found in previous
studies.\cite{PRBR,fujimori,kbmbnd,djsingh} It is evident that
spectral lineshape changes significantly with the change is photon
energy. The He {\scriptsize I} spectra are most surface sensitive
($\sim$ 80\%). The surface contribution in the $x$-ray photoemission
spectra is significantly small ($\sim$ 40\%). Thus, the different
spectral lineshape in Fig. 1(a) and 1(b) indicates that the surface
and bulk electronic structures are significantly different.

The extracted surface and bulk spectra are shown in Fig. 1(c) and
1(d). It is evident that the ultraviolet photoemission spectra
collected using thin films\cite{kkim} as well as the bulk samples in
the present study are very similar to the surface spectra obtained
from bulk samples. In the surface spectra and in the He {\scriptsize
I} spectra, the peak position of the most intense feature moves
towards higher binding energies with increase in $x$. While such
change may indicate a change in $U$ in the surface electronic
structure, it can have other origin too. For example, the
crystallographic symmetry at the surface is different from that in
the bulk leading to different crystal field effect, surface defects,
surface reconstructions {\it etc.} In fact it is believed that the
later effects are more reasonable in these systems.\cite{PRBR}

The bulk spectra exhibit finite density of states at the Fermi level
for all the $x$ values studied.\cite{PRB-Ravi} Most importantly, the
bulk spectra and/or the raw spectra obtained by $x$-ray
photoemission spectroscopy reveal that the energy position of the
electron correlation induced feature (lower Hubbard band) as marked
by vertical dashed lines in Fig. 1(a) and 1(c) moves towards the
Fermi level with the increase in Ti-dopant concentration in sharp
contrast to the conclusions of the concerned paper. Most
importantly, this observation is visible in the raw data of the
$x$-ray photoemission spectra and is independent of any data
analysis procedure. Thus, it is important to extract the surface and
bulk contributions from the experimental spectra to obtain realistic
representation of the bulk electronic structure. This also echoes
the view that the insulating nature observed at these compositions
are 'Anderson insulator' kind, which correspond to disorder induced
localized finite density of states at the Fermi level.

We now turn to the question of half-metallic phase, which is also
observed in earlier studies.\cite{kbm-srruti} The main aim of that
study was to grow half-metallic phase via Ti substitution induced
band narrowing. Since, the Ti 3$d$ levels has significantly
different eigen energies compared to the eigen energies of Ru 4$d$
levels and Ti 3$d$ band is almost completely empty appearing above
the Fermi level, the hopping of electrons from one Ru site to
another via Ti sites is smaller than that expected for Ru-O-Ru
hopping strength. Hence, Ti substitution leads to a significant
narrowing of the Ru 4$d$ bands. This in turn pulls the up spin band
completely below the Fermi level and only the down spin band
contributes at the Fermi level resulting to half metallicity. It is
already known that the electron correlation strength is
significantly weak in these systems.\cite{PRBR,fujimori} Thus, the
consideration of $U$ would slightly enhance the energy gap in the up
spin channel. The down spin channel will continue to contribute at
the Fermi level as finite intensity is observed in the experimental
bulk spectra. Thus, the conclusion of half metallic phase will not
be influenced significantly by the weak electron correlation in the
vicinity of $x$ = 0.5 compositions.

In summary, the surface and bulk electronic structure in
Ti-substituted SrRuO$_3$ are significantly different. The bulk
spectra in these ruthenates exhibit finite density of states at the
Fermi level for $x$ as high as 0.6 in SrRu$_{1-x}$Ti$_x$O$_3$. This
is consistent with the other experimental results indicating
Anderson insulating phase rather than the hard gap insulators. The
insulator to metal transition in these systems are primarily driven
by bandwidth, $W$ rather than $U$, and disorder plays an important
role in determining the electronic properties of these systems.


\begin{thebibliography}{99}

\bibitem{jkim} J. Kim, J.-Y. Kim, B.-G. Park, and S.-J. Oh, Phys. Rev. B
{\bf 73}, 235109 (2006).

\bibitem{abbate} M. Abbate, J.A. Guevara, S.L. Cuffini, Y.P. Mascarenhas, and E.
Morikawa, Eur. Phys. J. B {\bf 25}, 203 (2002).

\bibitem{kkim} K.W. Kim, J.S. Lee, T.W. Noh, S.R. Lee, and K. Char, Phys. Rev.
B {\bf 71}, 125104 (2005).

\bibitem{lin} P.-A. Lin, H.-T. Jeng, and C.-S. Hsue, Phys. Rev. B
{\bf 77}, 085118 (2008).

\bibitem{FujiRMP} M. Imada, A. fujimori, and Y. Tokura, Rev. Mod. Phys.
{\bf 70}, 1039 (1998).

\bibitem{GabiRMP}  A. Georges, G. Kotliar, W. Krauth, and M.J. Rozenberg, Rev. Mod.
Phys. {\bf 68}, 13 (1996).

\bibitem{PRBR} K. Maiti and R.S. Singh, Phys. Rev. B {\bf 71}, 161102(R) (2005).

\bibitem{fujimori} M. Takizawa, D. Toyota, H. Wadati, A. Chikamatsu, H. Kumigashira,
A. Fujimori, M. Oshima, Z. Fang, M. Lippmaa, M. Kawasaki, and H.
Koinuma, Phys. Rev. B {\bf 72}, 060404(R) (2005).

\bibitem{cao}  G. Cao, S. McCall, M. Shepard, J.E. Crow, and R.P. Guertin,
Phys. Rev. B {\bf 56}, 321 (1997).

\bibitem{LCVOPRL} K. Maiti, Priya Mahadevan, and D.D. Sarma, Phys.
Rev. Lett. {\bf 80}, 2885 (1998).

\bibitem{Y2Ir2O7} R.S. Singh, V.R.R. Medicherla, K. Maiti, and E.V.
Sampathkumaran, Phys. Rev. B {\bf 77}, 201102(R) (2008).

\bibitem{PRB-Ravi} K. Maiti, R.S. Singh, and V.R.R. Medicherla, Phys.
Rev. B {\bf 76}, 165128 (2007); {\it ibid.}, arXiv/0704.0327.

\bibitem{kbmbnd} K. Maiti, Phys. Rev. B {\bf 73}, 235110 (2006).

\bibitem{djsingh} David J. Singh, J. Appl. Phys. {\bf 79}, 4818
(1996).

\bibitem{kbm-srruti} K. Maiti, Phys. Rev. B {\bf 77}, 212407 (2008);
{\it ibid}, arXiv/0704.0321.

\end{thebibliography}
\end{document}